\documentclass[a4paper,11pt]{article}
\usepackage{url}
\usepackage{subfigure}
\usepackage{amssymb}
\usepackage{natbib}
\usepackage{graphicx}
\usepackage{a4wide}

\newcommand{\code}{\texttt}

\begin{document}

\title{Developing a Video Steganography Toolkit}

\author{James Ridgway \& Mike Stannett\\
  Department of Computer Science, University of Sheffield,\\
  Regent Court, 211 Portobello, Sheffield S1 4DP, United Kingdom\\[6pt]
  \url{james@james-ridgway.co.uk} \qquad \url{m.stannett@sheffield.ac.uk}
}

\maketitle

\begin{abstract}
Although techniques for separate image and audio steganography are widely known, relatively little has been described concerning the hiding of information within video streams (``video steganography''). In this paper we review the current state of the art in this field, and describe the key issues we have encountered in developing a practical video steganography system. A supporting video is also available online at \url{http://www.youtube.com/watch?v=YhnlHmZolRM}.
\end{abstract}

\textbf{Keywords.}
  Steganography; video manipulation; covert communications; information hiding; tool development.


\section{Introduction}

This paper reports the findings of a seven-month dissertation project carried out at Sheffield University, investigating steganographic techniques for hiding data in video files encoded using the popular H.264 format \citep{Ridgway13}. During the course of the project several key tools were developed, and these are described below together with experimental findings. All of the materials developed for the project can be accessed online at \url{http://www.steganosaur.us}. A supporting video is also available online at \url{http://www.youtube.com/watch?v=YhnlHmZolRM}.

\label{litsurvey}
\label{sec:common-techniques}

We begin by reviewing various existing approaches to digital steganography, before explaining in section \ref{section:development} the specific issues that need to be addressed when developing these techniques for video container files. In section \ref{sec:conclusion}, we describe our experimental findings, and highlight areas where further research and development might be beneficial.

\subsection{Background}
Whereas encryption seeks to make a message uninterpretable to unauthorised eavesdroppers, steganography attempts instead to make the very existence of the message unsuspected (the two techniques can of course be combined; see section \ref{sec:crypto-subsystem}). Steganography has a very long history: Herodotus explains how Histi{\ae}us, the `tyrant' of Miletus, who was then staying with his overlord (the Persian emperor, Darius I), had a message tattooed onto a slave's shaved head some 2500 years ago (499 BCE). Once the hair had grown back, the slave was sent to Miletus, where his nephew re-shaved the slave to find an instruction telling him to revolt against Darius \citep{Herodotus1}.

This idea, of hiding information covertly so that its presence is unsuspected even by eavesdroppers with access to the manipulated `container' (in this case, the slave), can of course be adapted to modern digital communications. Perhaps the simplest approach to digital steganography is to \emph{inject} data into redundant sections of a file. For example, because \code{EXE} files use an \emph{end of file} (EOF) marker, adding additional data to the end of the file doesn't affect executable behaviour. Other file types, e.g. WAV files, specify their intended size in a header \citep{Ibm04}, and additional data is again ignored. Although easy to implement, such injection techniques are extremely insecure, since direct analysis of the file can easily reveal the presence of unwarranted additional data.

\begin{figure}  
 \begin{center}  
  \fbox{\includegraphics*[natwidth=4in,natheight=1.42in]%
    {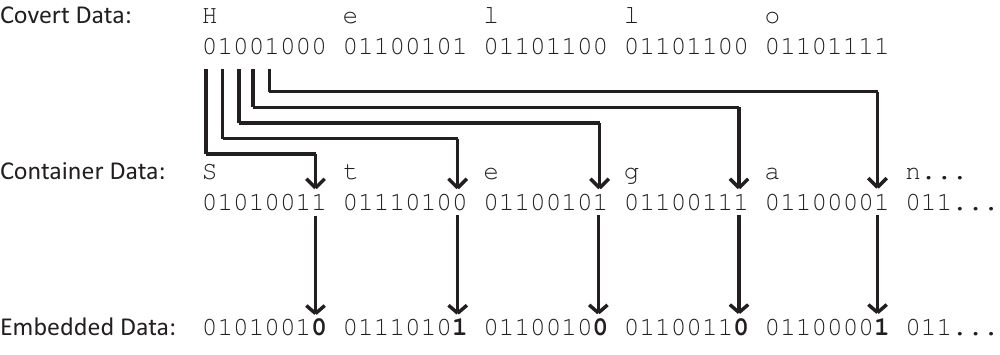}}
  \caption{\small LSB encoding of the word ``Hello'' inside an arbitrary container file.\label{fig:lsb1}}  
 \end{center}  
\end{figure}

In contrast, \emph{substitution} techniques embed data in those sections of the file that are -- relative to some appropriate metric -- of least relevance, without affecting overall file size. This avoids, e.g., the tell-tail size inflation associated with injection techniques, but the \emph{steganographic capacity} of the container is limited by the amount of `irrelevant' data present. A particularly common substitution technique for audio and image files is \emph{LSB manipulation} \citep{Johnson03,Cole03,Fridrich10} (Fig. \ref{fig:lsb1}), in which the least significant bit of each byte of an image, say, is manipulated so as to embed information without discernably changing the image as viewed on-screen. However, such techniques are inherently at odds with the lossy compression algorithms used by various digital encoding formats, since these specifically seek to disregard the same `irrelevant' segments of a file, i.e., they `throw away' precisely the segments where we want to hide our message. \emph{Transform-domain} techniques can sometimes overcome this problem (section \ref{transformdomain}).
%


\subsection{Video Steganography in the Literature}\label{lit:videosteg}

From a steganographic standpoint, video files have distinct advantages over stand-alone audio or image containers. In the first place, video files are typically much larger than other container files, and have far greater steganographic capacity. But in addition, video modification is also significantly harder for humans to detect than stand-alone image manipulation, because each video frame is only visible for a fraction of a second during normal playback, and moreover, video frames rarely include sharply focused images \citep{Al-Frajat10}.

Surprisingly, however, the number of articles addressing video steganography appears to be rather limited, and those that exist in the literature generally give high level descriptions with only limited lower-level detail. Of those authors who specifically address the topic, Node et al. (\citeyear{Noda04}) suggest a \emph{bit plane complexity segmentation} technique that hides data in wavelet-compressed video, and Jalab et al. (\citeyear{Jalab09}) give a related method for embedding data in MPEG video. Eltahir et al. (\citeyear{Eltahir09}) discuss LSB manipulation of video, but do not address security. Finally, \cite{Singh2} specifically address the hiding of an image in a video, and while their method centres around LSB manipulation, this is one of the few papers that exploits the multi-dimensional aspect of a video as a container file; they claim, moreover, that the proposed technique is ``very useful in sending sensitive information securely'' but unfortunately they provide little supporting evidence for this claim, or for the effectiveness of the proposed technique.

\subsection{Generation Techniques}
\label{generation-techniques}

Generation techniques involve generating a bespoke file from scratch by exploiting shared characteristics of the agents involved. For example, if Alice and Bob are both car enthusiasts, it is unlikely that the exchange of images showing pictures of the latest models will arouse much suspicion. We could therefore exchange a message covertly by creating an image of a race (say) in which the positions of the cars, spectators, or any other agreed component are used to encode the required information. Doing so may, of course, be time consuming, but this need not be an issue unless the information needs to be encoded and transmitted in real time. Such techniques have, moreover, a unique advantage over other steganographic methods, in that there is no underlying container file against which the transmitted file can be compared for steganalytic purposes. Reported research into generation techniques is, however, extremely limited, perhaps because the message construction process, with its heavy dependence on the shared interests of the specific agents involved, is necessarily ad hoc.

\subsection{Transform Techniques}
\label{transformdomain}
\label{lit:transform}

\emph{Discrete cosine transform} (DCT) techniques are often used with compressed image files, and these can be applied, to some extent, to individual images within certain video streams provided the frames to be manipulated are chosen appropriately. Informally, the discrete cosine transform takes image descriptions given in terms of pixel intensities, and re-expresses them in terms of frequencies, storing coefficients losslessly; these techniques are commonly used with, e.g., JPEGs \citep{Anderson96, Zhao95, Ruanaidh96}. Many existing steganographic systems make use of DCT coefficients, including the \emph{F5} \citep{Westfeld01} and \emph{Outguess} \citep{Provos01} algorithms, together with other \textit{model-based} \citep{Sallee03,Sallee05}, \textit{modified matrix} \citep{Kim06} and \textit{perturbed quantization} methods \citep{Fridrich05}.

Of more significance for our purposes, Prabhakaran and Shanthi (\citeyear{Prabhakaran12}) describe a hybrid crypto-steganography method, which \cite{Shanableh12} extends by encoding data in the motion vector and quantisation scales (section \ref{lit:videoencoding}); their technique increases the steganographic capacity of the file, but is generally limited to raw video. Fang and Chang (\citeyear{Fang06}) focus on modifying the motion vectors of fast-moving objects (since such changes are relatively undetectable). In contrast, \cite{Aly11} examines macroblocks to determine which are most suitable for LSB embedding; both papers report that the resultant video quality remains good, but this is not quantified.

\subsection{Video Encoding}\label{lit:videoencoding}

In this section we briefly describe the structure of a typical video file. Different types of video frame serve different purposes, and it is essential for steganographic purposes that only certain kinds of frame are manipulated; attempting to hide data within the wrong frames typically causes the covert message to become garbled during re-extraction. We consider these practical issues in more detail in section \ref{section:development} below.

\subsubsection{Coding Concepts}
An encoder (compressor) and decoder (decompressor) forming a complementary pair is known as a \emph{codec} (en\emph{co}der/\emph{dec}oder). The encoder is used to store or transmit video by converting the original raw video format to an alternative (compressed) representation. The decoder converts the compressed form back to the original video.

In general terms, a video encoder implements three main components: a \emph{temporal model}, a \emph{spatial model} and an \emph{entropy encoder}. The temporal model reads in a sequence of video frames and attempts to reduce redundancy by identifying similarities between neighbouring frames -- this analysis usually involves computing a prediction of the current video frame. With H.264 the prediction can be computed from multiple previous or future frames. The prediction is improved by means of compensation for differences between the frames -- this is known as motion compensation prediction. The temporal model outputs a residual frame and a set of \emph{motion vectors}. The residual frame is computed by subtracting the prediction from the current frame, and motion vectors are used to describe how the motion was compensated.

The residual frame from the temporal model is then fed into the spatial model. This step is again concerned with removing redundancy, but in this case by analysing neighbouring samples within the frame itself. Spatial reduction in H.264 is achieved by applying a transform followed by a \emph{quantisation} process. Quantisation is the process of scaling down the range of symbols that are used in a representation. For instance, the DCT transform produces a matrix of coefficients whose values may range between $-223$ and $150$, but after quantisation, these values may only range between $10$ and $130$. This reduced range means that fewer bits are needed to code the representation than the original range, and this can lead to significant (but lossy) compression (section \ref{sec:compression}). Quantisation parameters for multimedia formats are chosen based on how individual components affect the average human perception \citep{Mukhopadhyay1}. The transform step produces a set of transform coefficients which are then quantised, removing insignificant values, and returning the quantised transform coefficients as the output of the spatial model.

Finally, the entropy encoder produces an encoded output from the results of the spatial and temporal models. It processes the motion vectors from the temporal model and the coefficients from the spatial model to produce a compressed bit stream consisting of motion vectors, residual transform coefficients and header information.

Although the quantisation stage causes a loss of information, this process is roughly reversible, and the decoder mechanism essentially `works in reverse' to retrieve the original video. Nonetheless, the output produced by the decoder mechanism will only ever (in the case of H.264) be an approximation to the original input because of the quantisation stages.

\subsubsection{Temporal Model}
The residual frame produced by the temporal model is produced by subtracting the predicted frame from the actual video frame, and its size is dependent on the accuracy of the prediction process -- the smaller the residual frame, the fewer bits needed to code it. Prediction accuracy can be improved by calculating and propagating compensation for motion from the reference frame(s) through to the current frame.

Motion compensation can significantly improve prediction calculations because two successive video frames are usually highly correlated, because most of the information captured in successive residual frames relates to the movement of objects in an essentially static scene. These changes directly correspond to the movement of pixels between frames, a feature known as \emph{optical flow} \citep{Ahmad1}.

In theory, knowing the optical flow allows us to predict the majority of the pixels in the current frame, simply by displacing pixels in the preceding frame as required. Unfortunately, this is a very computationally intensive process, as each pixel will have to be transformed, and each frame decoded, on a pixel-by-pixel basis using the optical flow vectors. Whilst workable in theory, this would result in a large amount of residual data, which is at odds with the desirability of a compact residual frame.

\subsubsection{Macroblock Motion Estimation}
A macroblock is typically a $16 \times 16$ pixel block of the current frame, although in the wider context of block-based motion estimation other suitably-sized $N \times M$ samples might be used. Macroblocks are used by a variety of codecs including MPEG-1, MPEG-2, H.261, H.263 and H.264.

Macroblock motion estimation starts by dividing frames into macroblocks. Each macroblock is taken in turn, and a previously selected `reference frame' is searched for a matching macroblock. Macroblocks from the reference frame are paired with macroblocks in the current frame by choosing a candidate block that minimises the difference between the macroblock in the current frame and itself -- this process provides a \emph{residual block}. Finally, the residual block is encoded and stored, together with the associated motion vector.

Using a $16 \times 16$ size macroblock can cause some problems with certain motions and object outlines. If a macroblock and its matching macroblock differ greatly, the number of bits required for the encoding increases and inflates the bit-rate. This issue can be addressed by decomposing a macroblock into smaller $8 \times 8$, or even $4 \times 4$ macroblock size, but this results in a larger number of blocks, which can be disadvantageous. The H.264 codec overcomes this problem to some extent by adopting an `adaptive' block size approach.


\subsection{Compression}
\label{sec:compression}
An image in a video stream can be thought of as a function which maps each point of a 2D spatial domain to a three-dimensional RGB colour vector. Consequently, if we were to store a single 1920 x 1080 image in raw format, just over 2 million RGB triples would need to be stored. This is a substantial amount of information to store for a single image. Given that videos typically run at between 24 and 30 frames per second, a single second of video footage at 1920 x 1080 resolution would require the storage of around 50--60 million RGB triples. Storing and/or streaming this data in a raw, uncompressed format is consequently impractical for most situations, and as a result image and video formats typically use an alternative, compressed, representation.

\subsubsection{Compression Considerations}\label{lit:motvec}
Data compression inevitably involves trade-offs between computational costs, storage requirements, and accuracy of representation. For images and video, an approximation of the original source if often sufficient, which means that lossy compression schemes can be used, but if the given application requires complete accuracy then a lossless representation must be used. 

In section \ref{lit:transform} we saw how coefficient-based transforms such as DCT can be used to represent an image, and these techniques can be used to assist the image compression process. However, simply compressing each frame in turn is inefficient, since it fails to take into account the temporal cohesion between consecutive frames. Examining the correlation between consecutive frames typically allows a more concise representation to be used \citep{Mukhopadhyay1, Richardson1, Gall1}.

A common approach is to use a GOP (``group of pictures'') structure. In a GOP, a reference frame is chosen, which is called an \textit{Intra Frame} or \textit{I-Frame}. Other frames are then predicted from this, where predictions are represented as changes (deltas) from the preceding frame -- these are known as \textit{P-Frames}. Some frames are predicted using both the preceding and subsequent neighbours, and these are known as bi-directionally predicted frames or \textit{B-Frames} \citep{Mukhopadhyay1, Richardson1}.

\section{Development Issues}
\label{section:development}

Given the experimental focus of this work, we adopted an `agile' methodology and developed various auxiliary tools for embedding messages in audio and image files. In order to ensure usability, we designed the system as a user-friendly GUI, interacting with a lower-level suite of tools housing the core steganographic logic. We had intended using the same programming language throughout, but this proved infeasible, so different parts of the system had to be constructed using different programming languages.

In particular, there was insufficient time to develop a complete video coding tool from scratch, so we adopted third-party libraries. We began with Xuggler,\footnote{\url{http://www.xuggle.com/xuggler}} a Java wrapper for FFmpeg,\footnote{\url{http://ffmpeg.org/}} but unfortunately Xuggler proved insufficiently flexible, and we found it necessary to work directly with the FFmpeg library using code written in C. Even so we needed to re-implement part of the FFmpeg library -- we are grateful to Michael Niedermayer, one of FFmpeg's developers, for vital feedback at this time (personal communication).

Since we were already using C for the low-level video coding tools, we considered using it for the GUI as well, but this would have required using the GTK+ library,\footnote{\url{http://www.gtk.org/}} which has poor Mac OS integration. Moreover, good GUIs should be multithreaded to ensure the display updates regularly, but C has no uniform cross-platform API for managing threads. We therefore re-adopted Java and Xuggler for GUI development (Java includes GUI design packages, while Xuggler can be used for video playback).

\subsection{Choice of codec}
For development purposes we limited our choice of codecs to those supported by FFmpeg, focussing eventually on `H.264', since this is the most common video codec used by modern cameras \citep{Pcworld} and for online videos \citep{Techcrunch}.

\subsection{Transcode Mechanism}
Video transcoding involves demultiplexing the original input file to distinguish audio from video data. The separate streams are then processed independently and re-encoded back into the output file. Our transcoder uses FFmpeg's \code{avcodec} and \code{avformat} libraries. The input is scanned for audio and video streams, and relevant codecs are loaded into memory. A header is written to the output file, and we iterate over input data packets. Finally, a suitable footer is appended.

Though simple in theoretical terms, decoding input packets proved somewhat problematic in practice. Frames produced by FFmpeg's decoder methods \code{avcodec\allowbreak\_decode\allowbreak\_video2} and \code{avcodec\allowbreak\_decode\allowbreak\_audio4} cannot be parsed directly without preparation, because they lack appropriate timestamps.\footnote{A timestamp mechanism is used to synchronise different streams in a video file.} Failure to set these timestamps results in either no video image, or else a lack of synchronisation between audio and video streams. Frames decoded by \code{avcodec\allowbreak\_decode\allowbreak\_video2} also contained data that interfered with the encoding, causing the image to be heavily pixelated. We eventually solved this problem by copying raw image data to a new \code{AVFrame} instance, which allowed us to carry out the requisite pre-parsing preparation.

\subsubsection{Data encoding/decoding}
\begin{figure}
 \centering
 \subfigure[Main process]{
   \includegraphics*[natwidth=379pt,natheight=823pt,width=.4\textwidth]%
     {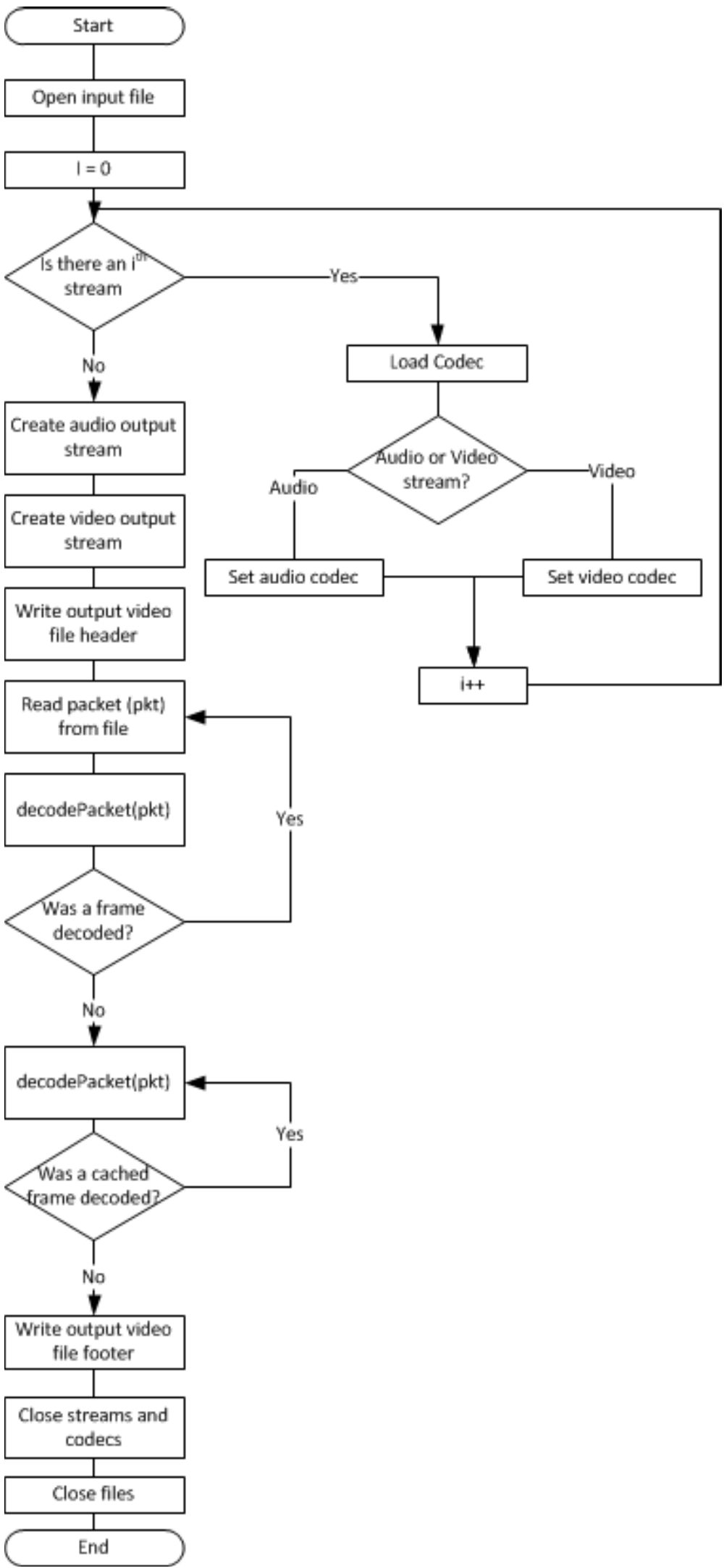}
   \label{fig:transcode_flow_chart}
 }
 \subfigure[\code{decodePacket} subprocess]{   
   \includegraphics*[natwidth=323pt,natheight=483pt,width=.35\textwidth]%
     {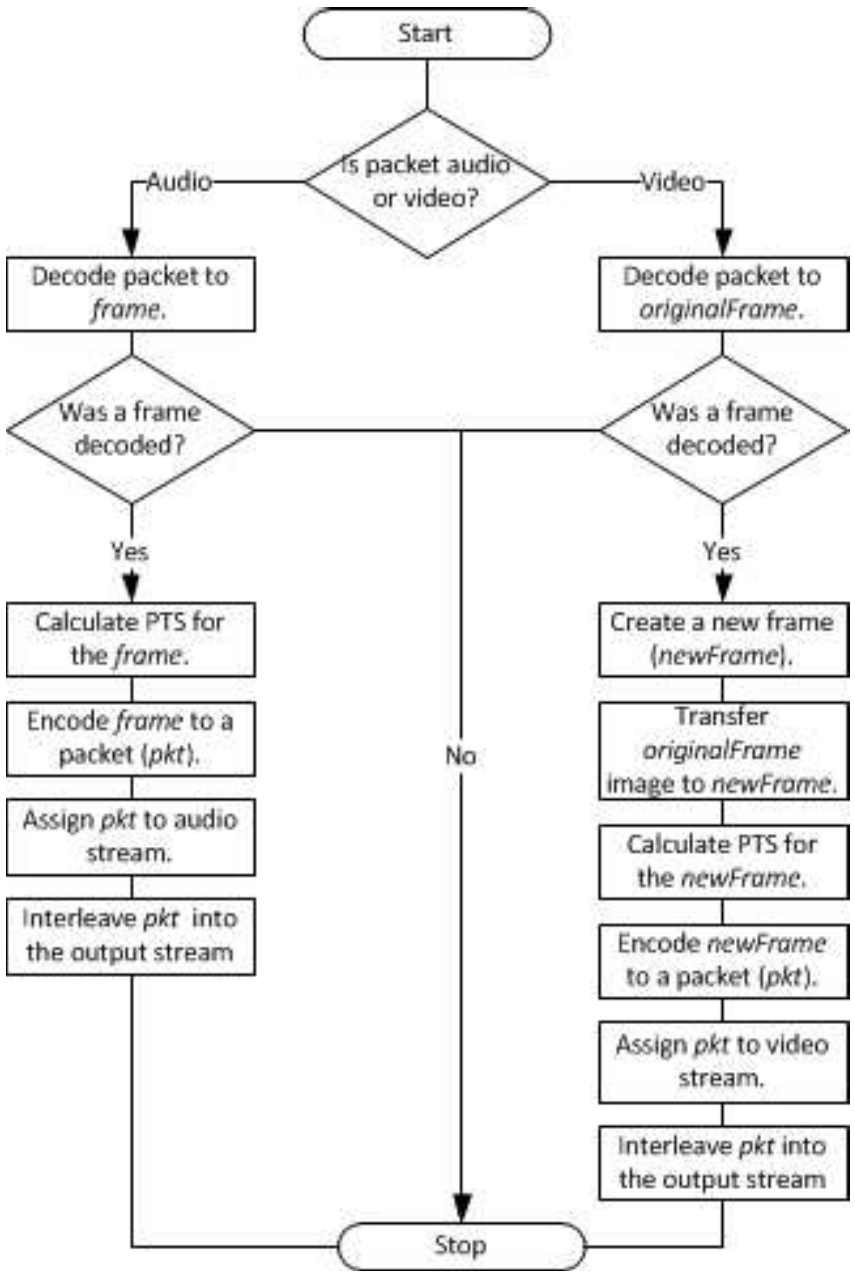}
   \label{fig:transcode_decode_packet_flow_chart}
 }
 \caption{Transcode process flow charts}
 \label{fig:transcoding}
\end{figure}

\begin{figure}
 \centering
 \subfigure[encoder]{
 \includegraphics*[natwidth=335pt,natheight=512pt,width=.35\textwidth]{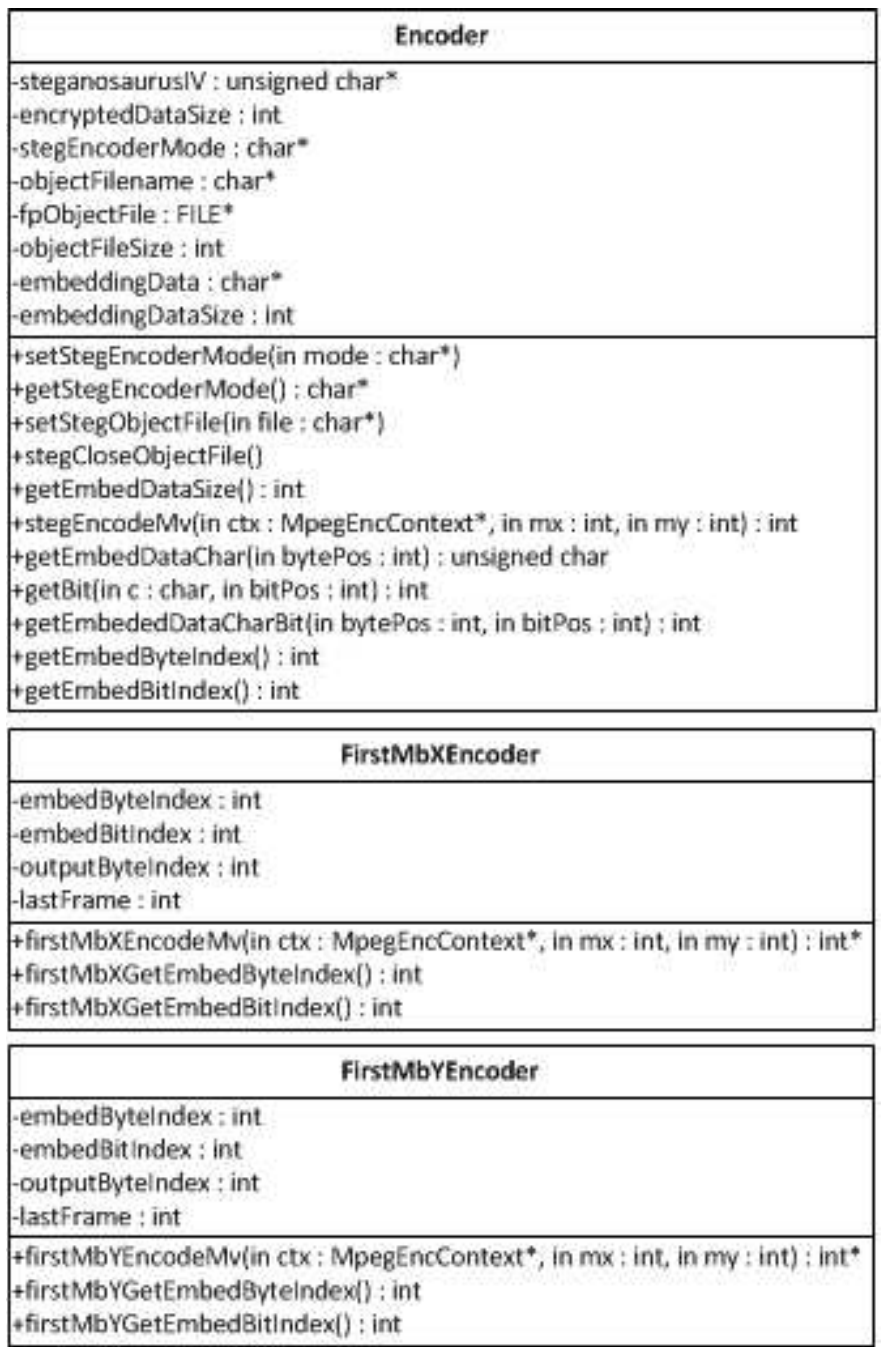}
 \label{fig:uml_encoder}
 }
 \subfigure[decoder]{
 \includegraphics*[natwidth=432pt,natheight=359pt,width=.45\textwidth]{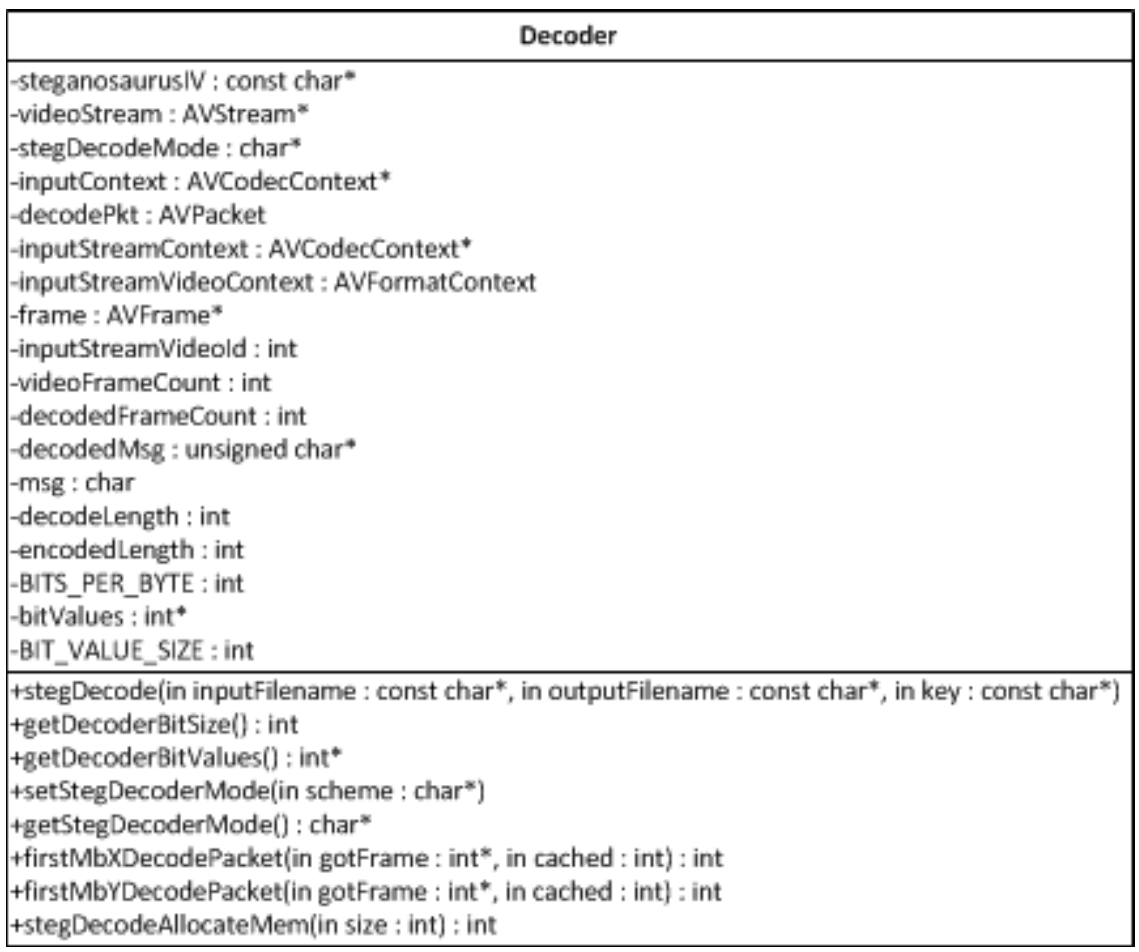}
 \label{fig:uml_decoder}
 }
 \caption{Encoder and decoder architecture overviews}
 \label{fig:uml_encoder_decoder}
\end{figure}

Data encoding and decoding was performed by modifying motion vectors using a callback (figure \ref{fig:transcoding}). This approach makes it easy to implement additional steganographic schemes: the motion vector and frame number are passed to our bespoke callback method, \code{stegEncodeMv} (invoked from inside the \code{avcodec} library), which chains a callback to the relevant encoder modules, as determined by the current \code{getStegEncoderMode} (figure \ref{fig:uml_encoder_decoder}). The decoding process is somewhat simpler -- it uses a single \code{Decoder} component (figure \ref{fig:uml_decoder}), iterating the relevant decoding method over the packets of the specified video file.

%
%

\subsection{Steganographic encoding/decoding}
We developed a process to modify the \code{AVFrame} parsed to the \code{avcodec\allowbreak\_encode\allowbreak\_video2} method for coding into a compressed packet, but despite our best efforts the alterations made to \code{motion\allowbreak\_val} were not reflected in the output. After dissecting the 850,000+ lines of code comprising the FFmpeg codebase, we deduced that adjusting the behaviour of \code{ff\allowbreak\_estimate\allowbreak\_p\allowbreak\_frame\allowbreak\_motion} in \code{libavcodec/motion\allowbreak\_est.c} would let us manipulate the motion vectors of macroblocks in P-Frames.

Initially, our encoder modified vectors using a bit mask, but tests showed that only around 30 characters could be embedded, and roughly 50\% of embedded bits would `flip', causing inaccurate decoding of the hidden message. Using different bit masks provided little improvement.

Eventually, we discovered that implementing our callback in the \code{ff\allowbreak\_estimate\allowbreak\_p\allowbreak\_frame\allowbreak\_motion} method of \code{libavcodec/motion\allowbreak\_est.c} caused data to be embedded prior to quantisation, on occasion obliterating our changes. We therefore moved our callback to the \code{encode\_mb\_internal} method of \code{libavcodec/\allowbreak{}mpegvideo\allowbreak\_enc.c}. Further analysis revealed that certain values returned by the decoder varied seemingly at random. We deduced that the vector coding process performs an additional right-shift operation, and therefore introduced a shifted encoding mask. The problem was fully resolved when we discovered that certain macroblocks were flagged as having no motion vector. After moving the encoder callback, compensating for bit shifts and avoiding vector-less frames our transcoding mechanism was finally capable of embedding data in video.

The decoder is far simpler, and operates by parsing \code{AVPacket}s to \code{avcodec\allowbreak\_decode\allowbreak\_video2} until a complete \code{AVFrame} has been returned.

\subsection{Cryptography Subsystem}
\label{sec:crypto-subsystem}
\begin{figure}
 \centering
 \subfigure[original image]{
  \includegraphics*[keepaspectratio,height=1.2in]{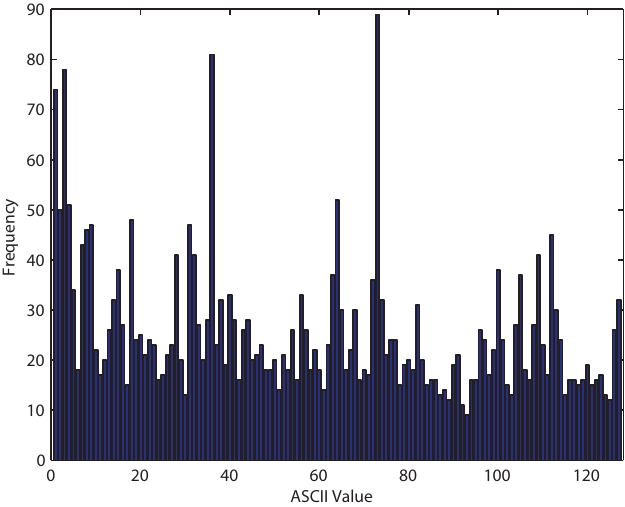}
  \label{fig:lsb2}
 }
 \subfigure[unencrypted embedding]{
  \includegraphics*[keepaspectratio,height=1.2in]{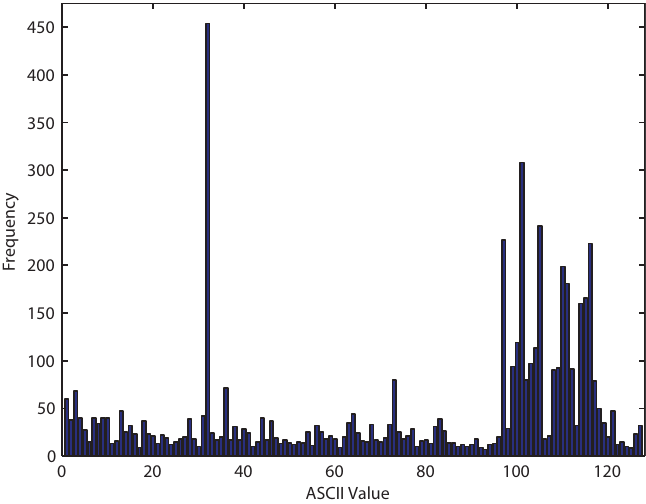}
  \label{fig:lsb3}
 }
 \subfigure[encrypted embedding]{
  \includegraphics*[keepaspectratio,height=1.2in]{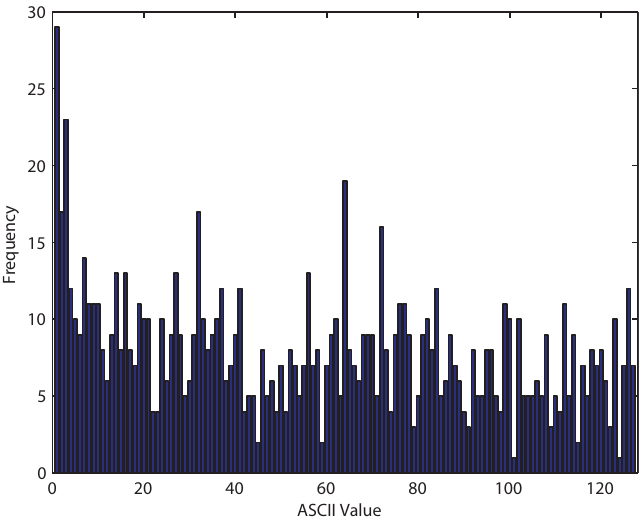}
  \label{fig:lsb4}  
 }
 \caption{ASCII distributions from the LSB string of a PNG image (a) before and (b) after unencrypted data is embedded; (c) the obfuscating effect of encrypting data prior to embedding.}
 \label{fig:lsb}
\end{figure}

It was not feasible, given the lifespan of the project, to determine whether our system exhibits longterm security. We therefore incorporated well-established cryptographic methods to ensure a minimum level of communications security. Our system fully implements the AES cryptosystem \citep{AES}, and comprehensive unit testing was employed to ensure that our implementation conforms to the relevant FIPS-197 specification \citep{FIPS197}. 
While encryption and steganography serve fundamentally different purposes, encryption can nonetheless be used \emph{with} steganography to obfuscate embedded ASCII messages: figure \ref{fig:lsb3} shows a significant increase in frequency for key ASCII values (especially spaces and alphabetic characters), but this tell-tale signature is removed by encrypting prior to embedding (figure \ref{fig:lsb4}).
%

\section{Evaluation and Conclusion}
\label{sec:conclusion}

Our initial intention was to develop numerous schemes for embedding and extracting data within video streams. This was overly ambitious, as we significantly underestimated the complexities associated with manipulating video. We nonetheless managed to develop two different methods for embedding and re-extracting data from video. Our final solution uses motion vector based approaches. 

This project also proved substantially more experimental than initially predicted. Our initial literature survey and preliminary research provided surprisingly little insight into how to design and implement a practical steganographic system. We therefore adopted an agile philosophy, and as the project progressed numerous design and implementation changes were undertaken and important lessons were learned. Although time constraints prevented us exploring more detailed steganographic schemes, our tools can take an object file in any format, apply AES encryption, and embed the data so that the resultant video is indistinguishable from the original and is capable of normal video playback. As part of testing and evaluation we made binary executables publicly available\footnote{\url{http://steganosaur.us/download}}.

\subsection{Technical limitations}

Our main goal was to develop and research techniques for embedding data using motion vector based techniques. While this has ultimately proven successful, we encountered notable setbacks. We thought it would be possible to embed data in motion vectors of any P- or B- frame, but discovered in some cases that a macroblock can be coded as having \emph{no} motion vector (as opposed to one of zero magnitude). This is an important distinction that needs to be considered carefully when choosing the specific frames in which to embed covert data.

Another limitation is our inability to determine steganographic capacity in advance. Motion vectors describe the spatial translation of a pixel block between frames, whence modifying one frame will impact its neighbours. Moreover, the number of encodable macroblocks changes with the object to be embedded, and can also vary due to keyframe positions or GOP sizes. This sometimes changes B- or P- frames to I-frames which have no macroblocks, dramatically changing the number available. Whilst we could change our system to preserve I-frame positions, we found that regular GOP sizes provide better error correction.

Nonetheless, we have produced a system that achieves most of the goals we set (see \ref{app:images} for examples of some of our outputs). The project was considerably more complex than originally envisaged, but we have largely been able to identify and overcome the major hurdles we encountered.

\appendix
\section{Typical system outputs}\label{app:images}

\begin{figure}
  \centering
  \subfigure[original]{
  \includegraphics*[natwidth=640pt,natheight=320pt,width=.4\textwidth,keepaspectratio]{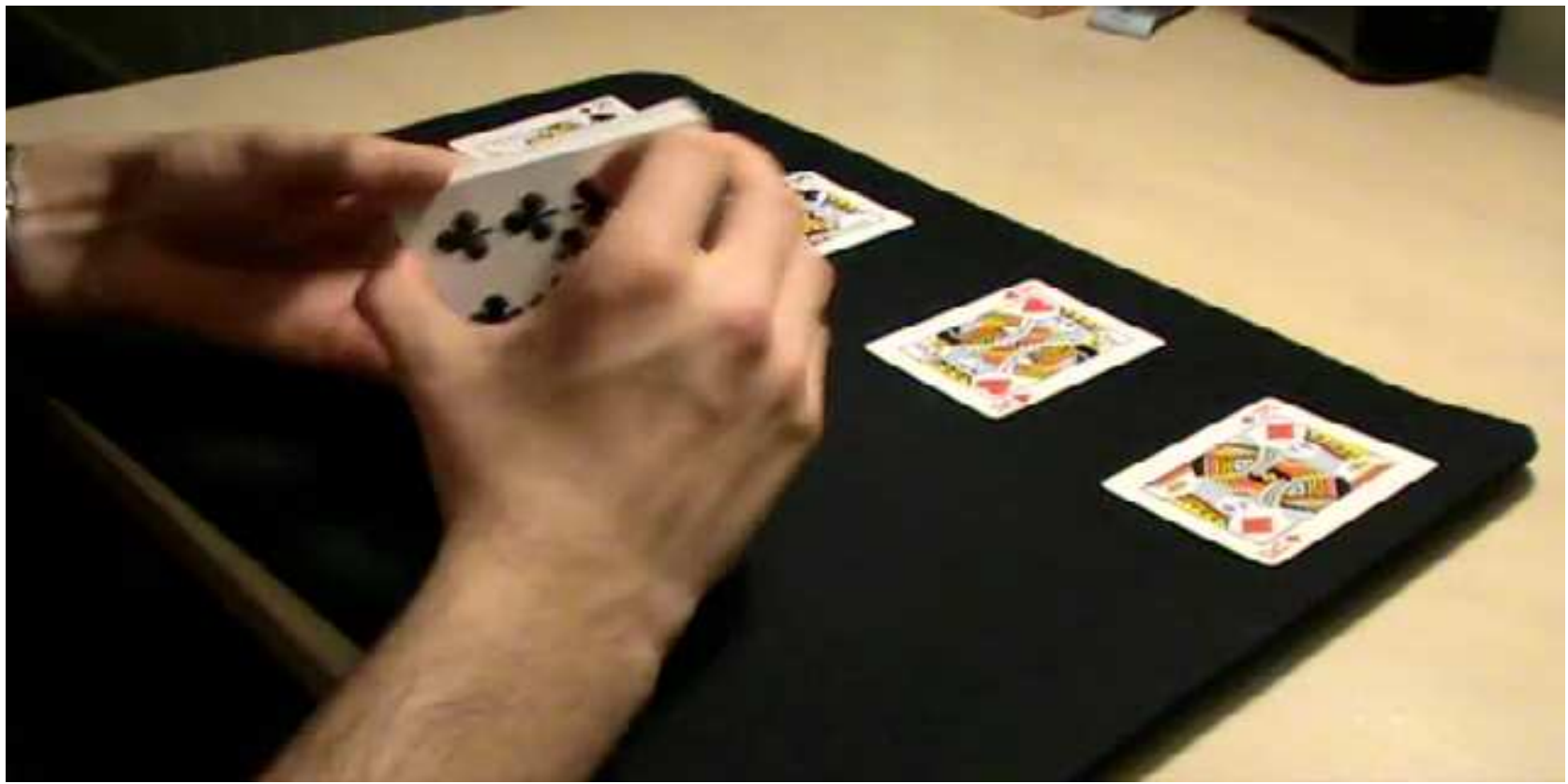}
  \label{fig:eval_ars_normal_f600}
  }
  \subfigure[inverted]{
  \includegraphics*[natwidth=640pt,natheight=320pt,width=.4\textwidth,keepaspectratio]{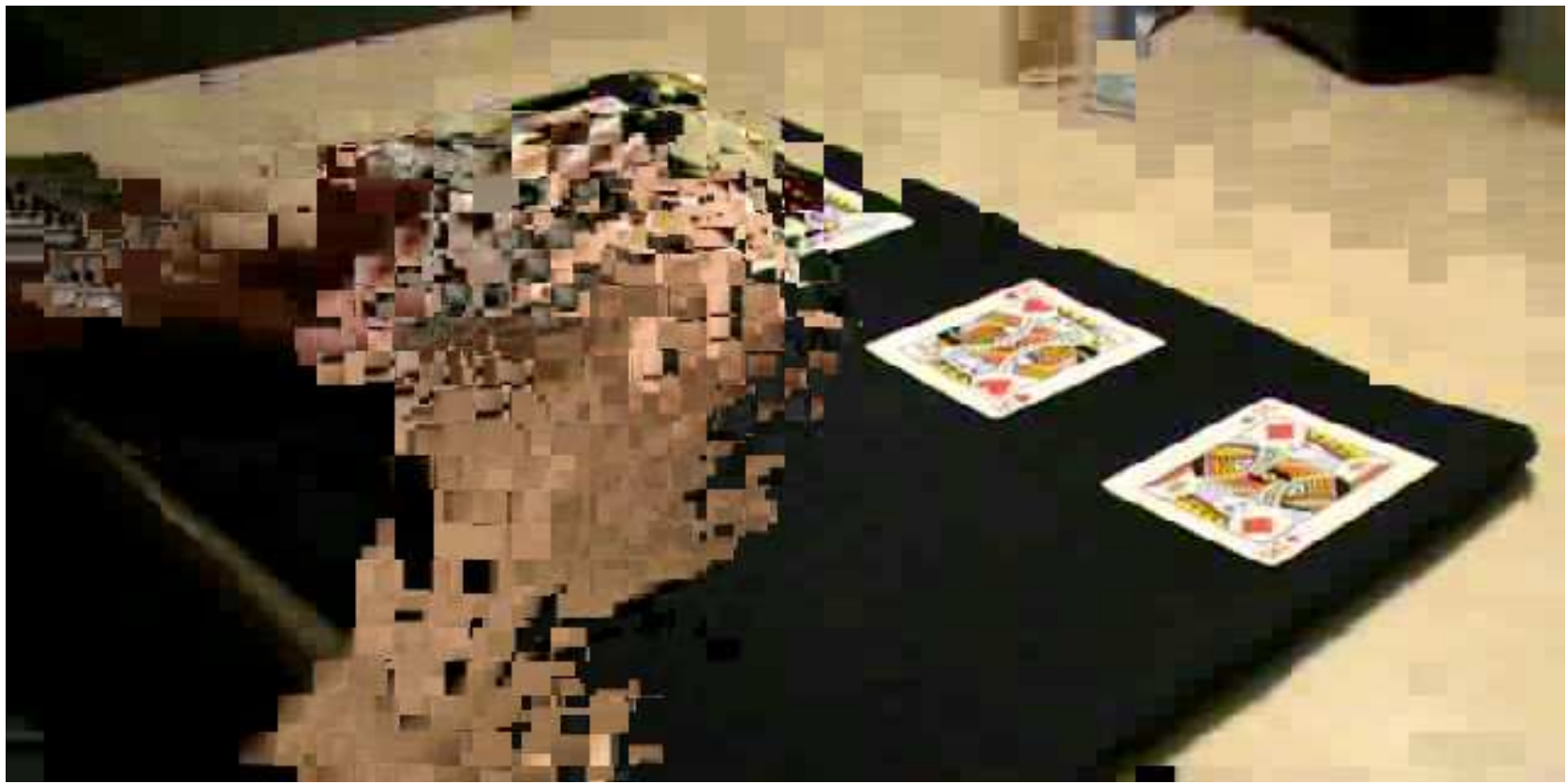}
  \label{fig:eval_ars_negative_f600}
  }
  \subfigure[\code{firstMbXEncodeMv}]{
  \includegraphics*[natwidth=640pt,natheight=320pt,,width=.4\textwidth,keepaspectratio]{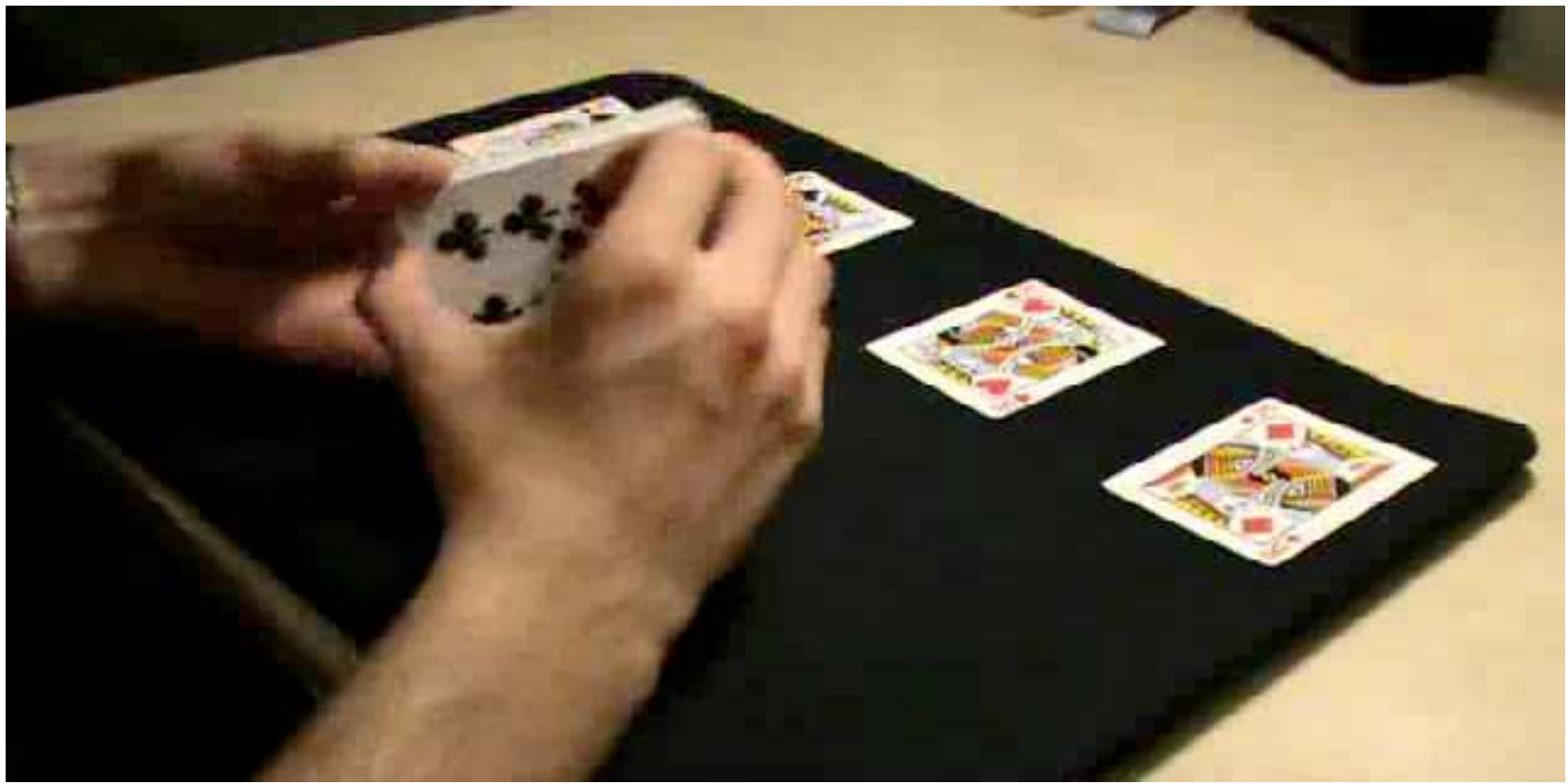}
  \label{fig:eval_ars_first_x_f600}
  }
  \subfigure[\code{firstMbYEncodeMv}]{
  \includegraphics*[natwidth=640pt,natheight=320pt,,width=.4\textwidth,keepaspectratio]{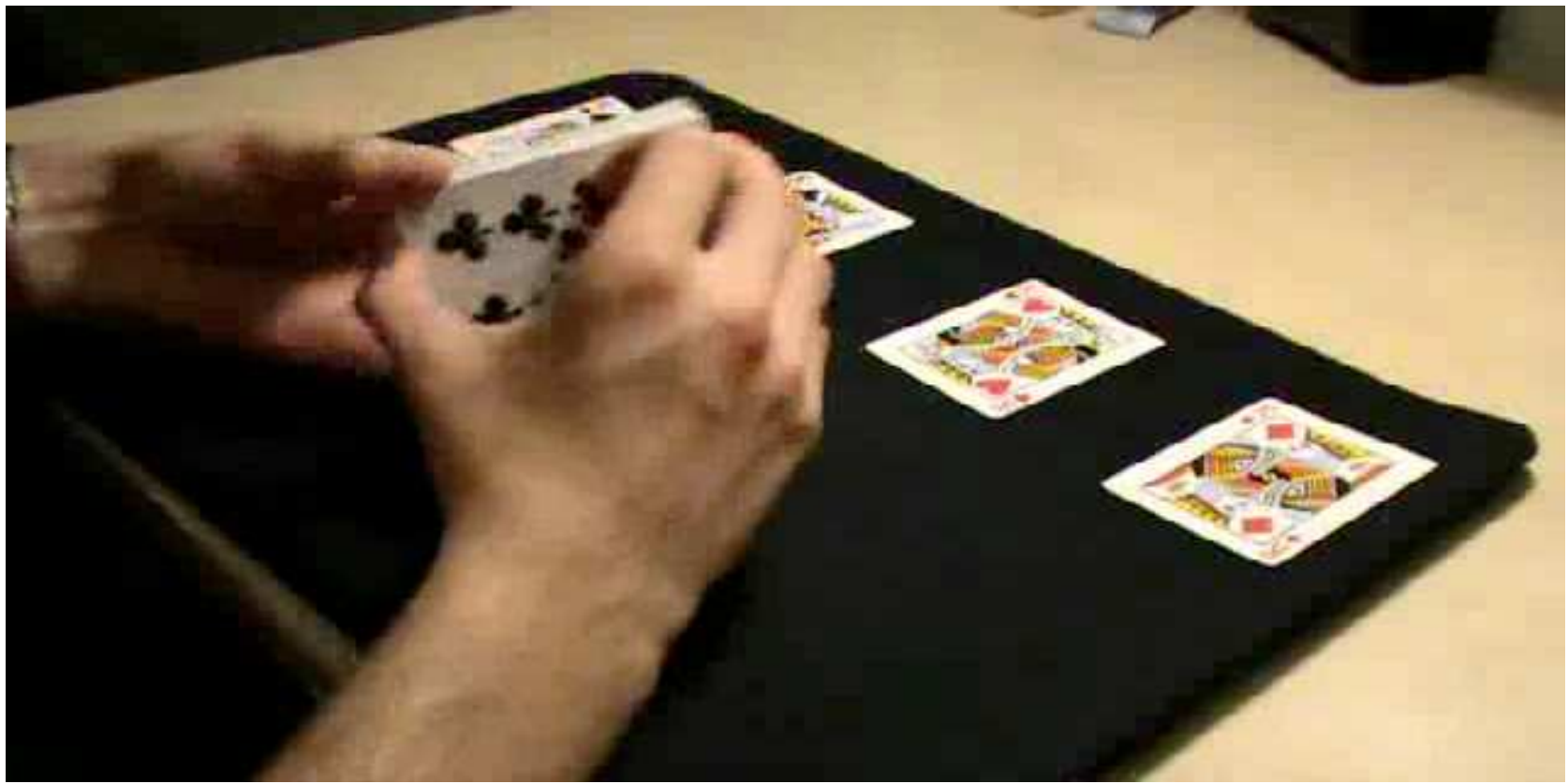}
  \label{fig:eval_ars_mby_f600}
  }
  \caption{The visible effects of various encodings: (a) original video frame; (b) inverted motion videos producing visible artefacts; (c, d) using our two encoding functions produces no obvious artefacts.}
  \label{fig:various-encodings}
\end{figure}

\begin{figure}
 \centering
 \includegraphics*[natwidth=640pt,natheight=486pt,width=.8\textwidth,keepaspectratio]{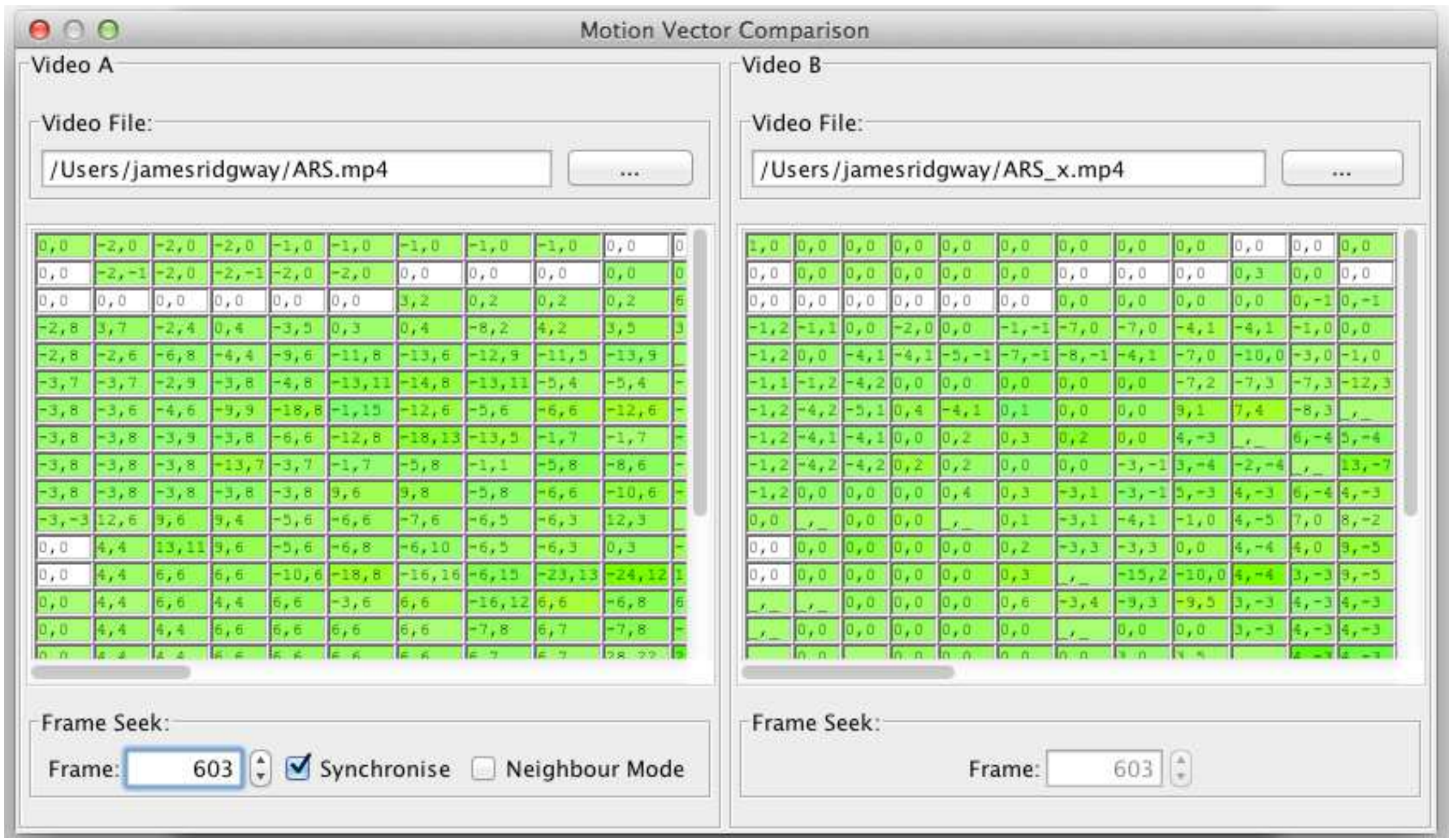}
 \caption{Motion vector comparison of figures \ref{fig:eval_ars_normal_f600} and \ref{fig:eval_ars_first_x_f600}.}
 \label{fig:eval_ars_f600_mvc}
\end{figure}

Figure \ref{fig:various-encodings} shows an unmodified frame (a) from one of our test videos, together with the effects of applying various modifications. In (b) we see how inappropriate encoding produces identifiable distortion. However, user testing showed that neither of the bespoke encoding methods \texttt{firstMbXEncodeMv} (c) or \texttt{firstMbYEncodeMv} (d) produced visibly identifiable artefacts. 
Figure \ref{fig:eval_ars_f600_mvc} shows the corresponding motion vectors in the original and \code{firstMbXEncodeMv} versions of a frame. The X-component of this frame was not modified, but there is nonetheless significant fluctuation in the motion vector values, due to the lossy nature of H.264/MPEG4 coding.
%

\bibliography{steg}

\end{document}